# Field-portable quantitative lensless microscopy based on translated speckle illumination and sub-sampled ptychographic phase retrieval

He Zhang,[1,2,4] Zichao Bian,[2,4] Shaowei Jiang,[2,4] Jian Liu,[1,5] Pengming Song,[3] and Guoan Zheng,[2,3,6]

[1]Ultra-Precision Optoelectronic Instrument Engineering Center, Harbin Institute of Technology, Harbin 150001, China
[2]Department of Biomedical Engineering, University of Connecticut, Storrs, CT, 06269, USA
[3]Department of Electrical and Computer Engineering, University of Connecticut, Storrs, CT, 06269, USA
[4]These authors contributed equally to this work
[5]Email: liujian@hit.edu.cn
[6]Email: guoan.zheng@uconn.edu

**We report a compact, cost-effective and field-portable lensless imaging platform for quantitative microscopy. In this platform, the object is placed on top of an image sensor chip without using any lens. We use a low-cost galvo scanner to rapidly scan an unknown laser speckle pattern on the object. To address the positioning repeatability and accuracy issues, we directly recover the positional shifts of the speckle pattern based on the phase correlation of the captured images. To bypass the resolution limit set by the imager pixel size, we employ a sub-sampled ptychographic phase retrieval process to recover the complex object. We validate our approach using a resolution target, a phase target, and a biological sample. Our results show that accurate, high-quality complex images can be obtained from a lensless dataset with as few as ~10 images. We also demonstrate the reported approach to achieve a 6.4 mm by 4.6 mm field of view and a half pitch resolution of 1 μm. The reported approach may provide a quantitative lensless imaging strategy for addressing point-of-care, global-health and telemedicine related challenges.**

http://dx.doi.org/

Achieving high spatial resolution and a wide field of view is the goal of many microscopy imaging systems. Designing a high numerical aperture (NA) lens with diffraction-limited performance over a large field of view remains a difficult task for conventional microscope platforms. Lensless microscopy, on the other hand, plays an especially prominent role in computational imaging platforms aimed at achieving high resolution and large field of view with compact optical arrangements [1-13]. In lensless microscopy platforms, if the sample is directly placed on top of the image sensor, the object intensity can be recovered via back projection [9, 10]. If there is a short propagation distance between the sample and the image sensor, the principle of digital in-line holography can be used to recover the complex object [1]. Multiple object heights [6, 14], multiple incident angles [7, 13], and multiple wavelengths [11, 12] can be used in the acquisition process to add diversity to the phase retrieval problem. Recently, Feng et.al. has also demonstrated the use of multiple illumination patterns for lensless phase retrieval [8].

In this letter, we report a compact, cost-effective and field-portable lensless imaging platform for quantitative microscopy. In this platform, we place the object at a short distance to the image sensor chip. We then use a low-cost galvo scanner to rapidly scan an unknown laser speckle pattern on the object. To address the positioning repeatability and accuracy issues, we directly recover the positional shifts of the speckle pattern based on the phase correlation of the captured images. To bypass the resolution limit set by the imager pixel size, we employ a sub-sampled ptychographic phase retrieval process to recover the complex object [15, 16]. We validate our approach using a resolution target, a phase target, and a biological sample. Our results show that accurate, high-quality complex images can be obtained from a lensless dataset with as few as ~10 images. We also demonstrate the reported approach to achieve a 6.4 mm by 4.6 mm field of view and a half pitch resolution of 1 μm. The reported approach may provide a quantitative lensless imaging strategy for addressing point-of-care, global-health and telemedicine related challenges.

Our prototype setup is shown in Figs. 1(a) and 1(b), where we use a 100-mW, 532-nm laser diode for sample illumination. A scotch tape is used as a diffuser and attached to the output port of the laser diode. The resulting light field on sample plane forms an unknown speckle pattern. We then use a low-cost galvo scanner (20Kpps Laser Galvo

Galvanometer, $116, Amazon) and a microcontroller (Arduino Due) to rapidly scan the unknown speckle pattern on the object. The images are captured by an image sensor (MT9J003 ON Semiconductor; 1.67 μm pixel size) placed ~0.5 mm under the object. The acquired images are then used to recover the complex object. Figure 1(c1) shows a raw captured image of a phase target under the speckle illumination. Figure 1(c2) shows the recovered phase of the object.

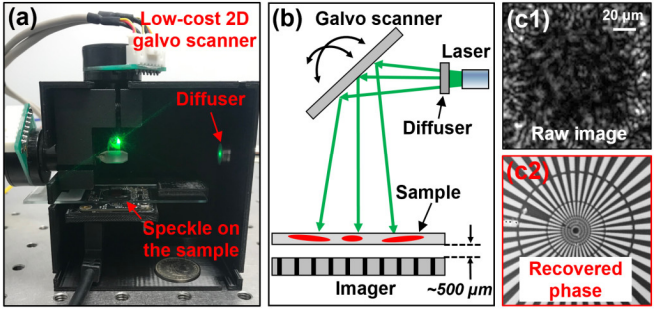

Fig. 1. The proposed lensless imaging platform based on translated speckle illumination and sub-pixel ptychographic phase retrieval. (a) The prototype setup, where we use a low-cost galvo scanner to rapidly translate an unknown speckle pattern on the object. The acquired images are used to recover both the complex object and the speckle pattern. (b) The optical scheme of (a). (c1) The acquired raw image of a phase target under the speckle illumination. (c2) The recovered phase of the phase target.

The forward imaging model of the captured images in this lensless imaging system can be described as

$$I_j(x,y) = \left|\left(O(x,y) \cdot P(x-x_j, y-y_j)\right) * PSF(d)\right|^2, \quad (1)$$

where $I_j$ is the $j^{th}$ intensity measurement ($j = 1,2,3 \dots, J$) with a size of $M \times M$ pixels, $O(x, y)$ is the complex object, $P(x, y)$ is the unknown speckle pattern, $(x_j, y_j)$ is the $j^{th}$ unknown positional shift of the speckle pattern, $PSF(d)$ is the point spread function (PSF) for free-space propagation over a distance $d$, and '*' stands for convolution operation. In our implementation, we use angular spectrum method to perform the free-space propagation in the Fourier domain [17].

Based on all captured images $I_j$ under the translated speckle illumination, the goal of our lensless imaging platform is to recover the complex object $O(x, y)$ and the speckle pattern $P(x, y)$. The recovery process is shown in Fig. 2. We first reorder the captured images to a random sequence. We then calculate the positional shift of the unknown speckle pattern using the phase correlation of the captured images. We initialize the amplitude of the unknown speckle pattern by shifting back all captured raw images, averaging them and padding the size from $M \times M$ to $3M \times 3M$ pixels in the frequency domain with zeros. Similarly, we initialize the amplitude of object by averaging all measurements and padding the size to $3M \times 3M$ pixels.

As shown in Fig. 2, for the $j^{th}$ captured image, we multiply the object with the shifted speckle pattern to obtain the exit wave $\varphi_j(x,y)$ at the object plane. We then propagate the exit wave $\varphi_j(x,y)$ to the detector plane with a distance $d$ and obtain $\psi_j(x,y)$. In a typical phase retrieval process, the amplitude of $\psi_j(x,y)$ will be replaced by the $j^{th}$ measurement $\sqrt{I_j}$. In our implementation, we perform sub-pixel updating in order to bypass the resolution limit set by the pixel size [18]. Lines 8-9 in Fig. 2 correspond to the sub-pixel updating process and line 10 corresponds to the Fourier magnitude projection process. The updated $\psi_j'(x_j, y_j)$ is then propagated back to the object plane to obtain $\varphi_j'(x_j, y_j)$. We use the regularized ptychographical iterative engine (rPIE) [19] to update the object $O$ and the speckle pattern $P_j$ as follows:

$$O = O + \frac{conj(P_j)(\varphi_j' - \varphi_j)}{(1-\alpha_{obj})|P_j|^2 + \alpha_{obj}|P_j|^2_{max}}, \quad (2)$$

$$P_j = P_j + \frac{conj(O)(\varphi_j' - \varphi_j)}{(1-\alpha_P)|O|^2 + \alpha_P|O|^2_{max}}, \quad (3)$$

, where 'conj' denotes conjugate, and $\alpha_{obj}$ and $\alpha_P$ are algorithm weights in rPIE. We also add Nesterov momentum to accelerate the convergence speed in our implementation [19]. The processing time for 100 raw images with 512 by 512 pixels each is ~25 seconds for 10 iterations using a Dell XPS 8930 desktop computer.

---

**Algorithm outline**

**Input**: Raw images $I_j$ ($j = 1,2,\cdots,J$) under translated speckle illumination
**Output**: High-resolution object $O(x,y)$ and the speckle pattern $P(x,y)$

1. Calculate the translated position $(x_j, y_j)$ of the speckle pattern
2. Initialize $O(x,y)$ and $P(x,y)$
3. **for** $n$= 1: $N$ (different iterations)
4.     **for** $j$= 1: $J$ (different captured images)
5.         $P_j(x,y) = P(x - x_j, y - y_j)$
6.         $\varphi_j(x,y) = O(x,y) \cdot P_j(x,y)$
7.         $\psi_j(x,y) = PSF(d) * \varphi_j(x,y)$
8.         $Intensity = |\psi_j(x,y)|^2$
9.         $Intensity^{update}(1:3:3M, 1:3:3M) = I_j$
10.        $\psi_j'(x,y) = \psi_j(x,y)/|\psi_j(x,y)| \cdot \sqrt{Intensity^{update}}$
11.        $\varphi_j'(x,y) = PSF(-d) * \psi_j'(x,y)$
12.        Update object using Eq. (2)
13.        Update speckle pattern using Eq. (3)
14.        $P(x,y)$ =Shift back $\left(P_j(x,y)\right)$
15.        Add Nesterov momentum
16.     **end**
17. **end**

---

Fig. 2. The recovery process of the lensless imaging platform, where the acquired images under translated speckle illumination are used to recover both the complex object and the speckle pattern. The positional shift of the speckle pattern is unknown at the beginning of this process.

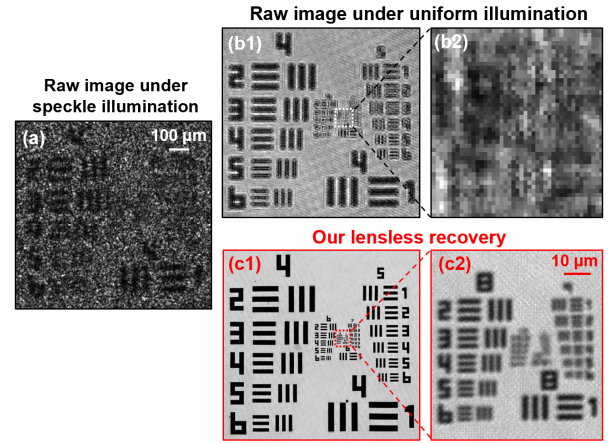

Fig. 3. The raw lensless image of a resolution target under speckle illumination (a), and uniform illumination (b). (c) Our lensless recovery based on 400 raw images.

We first validate the proposed lensless imaging scheme using a USAF resolution target in Fig. 3. To estimate the correct propagation distance $d$, we use a metric to quantify the convergence of the solution [20]. By maximizing this metric, we can then recover the correct propagation distance $d$. Figures 3(a) and 3(b) show the raw captured images under speckle and uniform illumination. Figure 3(c) shows our recovery, where we can resolve 0.98-μm linewidth from the group 9, element 1 of the resolution target. The positional shift of the speckle pattern is about 2-3 pixels in between adjacent acquisitions.

In the second experiment, we use a quantitative phase target (Benchmark QPT) as the object. Figure 4(a) shows the captured raw image under uniform illumination. Figure 4(b) shows the captured raw image under speckle illumination. Our lensless recovery is shown in Fig. 4(c). The line profile of the red dash circle in Fig. 4(c) is plotted in Fig. 4(d). The recovered phase is in a good agreement with the ground-truth height of the phase target. The recovered height profile of the phase target is shown in Fig. 4(e) for visualization.

In the fourth experiment, we test the proposed scheme using a blood smear sample. Figure 6 shows the recovered intensity and phase using different numbers of raw images. In this experiment, we use a pre-recovered speckle pattern in the reconstruction process. We can see that the individual blood cells can be clearly resolved using as few as 16 raw images.

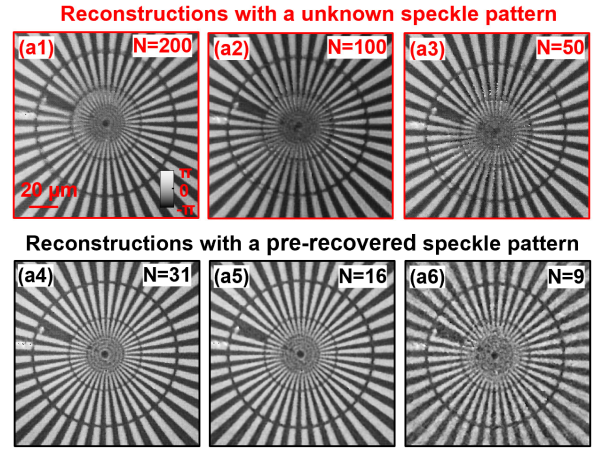

Fig. 5. (a) Reconstructions with an unknown speckle pattern and 200 (a1), 100 (a2), and 50 (a3) raw images. (b) Reconstructions with a pre-recovered speckle pattern and 31 (a1), 16 (a2), and 9 (a3) raw images.

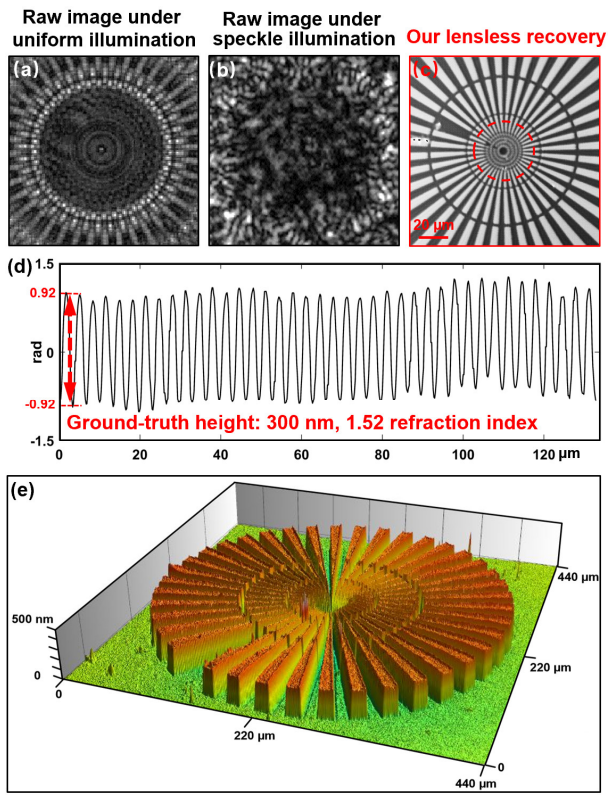

Fig. 4. Validating the quantitative nature of the reported scheme using a quantitative phase target. The raw lensless image under uniform illumination (a), and speckle illumination (b). (c) Our lensless recovery based on 400 raw images. (d) The line profile of the red circle in (c). (d) The recovered height profile of the phase target.

In the third experiment, we investigate the number of raw images needed for our recovery. Figure 5(a) show the recovered results using different numbers of translated positions with an unknown speckle pattern. The reconstruction quality increases as the number of captured raw images increases. In our implementation, we can also recover the unknown speckle in a calibration experiment. The pre-recovered speckle pattern can then be used to recover other unknown objects. Figure 5(b) shows the recovered phase target with the pre-recovered speckle pattern. In this case, we can reduce the number of translated positions to 9.

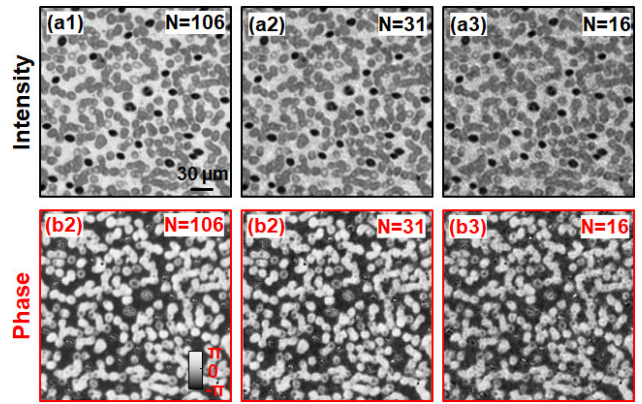

Fig. 6. Reconstructions of a blood smear target with different numbers of raw images. (a) The recovered intensity. (b) The recovered phase.

In many microscopy applications, it is important to achieve both high-resolution and wide field of view at the same time. Figure 7 shows the full field of view reconstruction of the blood smear sample. We use 400 raw images in the reconstruction process and the speckle pattern is treated as unknown. The imaging area is 6.4 mm by 4.6 mm, limited by the size of the image sensor. Figures 7(a) and 7(b) show the magnified intensity and phase of the two highlighted regions (a) and (b).

In summary, we report a compact, cost-effective and field-portable lensless imaging platform for quantitative microscopy. In our platform, we translate an unknown speckle pattern on the object using a low-cost galvo scanner. The captured images are used to recover both the complex object and the unknown speckle pattern. The proposed lensless imaging scheme shares its root with Talbot-illumination lensless imaging [8], near-field ptychography [15, 16], near-field Fourier ptychography [21], and translated speckle illumination microscopy [22, 23]. These schemes all use a translated non-uniform pattern for sample illumination. Different from these previous implementations, we employ a low-cost galvo scanner to rapidly scan the speckle to different positions, enabling a compact and field-portable configuration for quantitative lensless microscopy. The positional shifts of the speckle pattern are treated as unknown and we recover them based on the phase correlation of the acquired images. A sub-pixel updating scheme is further used to bypass the resolution limit set by the imager pixel size.

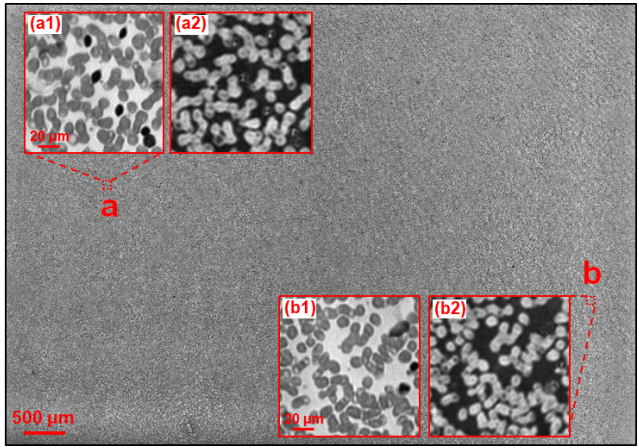

Fig. 7. The full field of view of the blood smear sample. Insets show the recovered intensity and phase images of regions (a) and (b).

The advantages of the reported platform can be summarized in threefold. First, we do not need to know the position of the speckle pattern. Therefore, the image acquisition process can be free-run. No synchronization, triggering or timing is needed. Second, the use of low-cost galvo scanner is advantageous compared to the regular multi-height implementations. The mechanical scanning time is neglectable in our implementation. Third, the reported platform provides the true quantitative contrast of the complex object. It may provide a solution for addressing point-of-care and telemedicine related challenges

There are several future directions of the reported scheme. First, we assume the object to be a 2D thin section in our implementation. For a thick object, multi-slice modeling can be used to recover the 3D information [24]. Second, the reported scheme can also be implemented at the detection path. In this case, we can place a diffuser in between the object and the detector. Two rotating glass slides can be then placed in between the object and the diffuser to shift the light wave. Implementing the scanning process at the detection path has the advantage of handling 3D thick object, similar to that in aperture-scanning Fourier ptychography [25]. Effort along this direction is ongoing. Third, we directly recover positional shifts of the speckle pattern using the phase correlation technique in the current implementation. The accuracy may be limited due to the influence of the object. A better strategy may be to update the positional shift in each iteration. Fourth, we can treat the speckle pattern as a special form of structured illumination. A speckle pattern with fine spatial feature may be able to significantly improve the final achievable resolution, similar to our recent demonstrations in fluorescence microscopy [23] and near-field Fourier ptychography [21].

**Funding.** This work is partially supported by National Science Foundation (1510077) and National Institute of Health (R21EB022378, R03EB022144).